\documentstyle[11pt,epsfig,here,mine]{article}
\begin{document}

\begin{center}
{\Large
{\it
Submitted to the Proceedings of 25th ICRC\\
(Durban, South Africa, July 28 - August 8, 1997)\\
}}
\end{center}

\vspace{-1.4cm}
\title{
SEARCH FOR NEUTRINOS FROM THE CORE OF THE EARTH
WITH THE BAIKAL UNDERWATER DETECTOR NT-36
}

\vspace{14mm}
\author{
{\large THE BAIKAL COLLABORATION:}\\[2mm]
V.A.BALKANOV$^2$, I.A.BELOLAPTIKOV$^7$, L.B.BEZRUKOV$^1$, B.A.BORISOVETS$^1$, 
N.M.BUDNEV$^2$, A.G.CHENSKY$^2$, I.A.DANILCHENKO$^1$, ZH.-A.M.DJILKIBAEV$^1$, 
V.I.DOBRYNIN$^2$, G.V.DOMOGATSKY$^1$, A.A.DOROSHENKO$^1$, S.V.FIALKOVSKY$^4$,
O.N.GAPONENKO$^2$, A.A.GARUS$^1$, S.B.IGNAT$'$EV$^3$, A.KARLE$^8$, 
A.M.KLABUKOV$^1$, A.I.KLIMOV$^6$, S.I.KLIMUSHIN$^1$, A.P.KOSHECHKIN$^1$, 
V.F.KULEPOV$^4$, L.A.KUZMICHEV$^3$, B.K.LUBSANDORZHIEV$^1$, T.MIKOLAJSKI$^8$,
M.B.MILENIN$^4$, R.R.MIRGAZOV$^2$, A.V.MOROZ$^2$, N.I.MOSEIKO$^3$, 
S.A.NIKIFOROV$^2$, E.A.OSIPOVA$^3$, A.I.PANFILOV$^1$, YU.V.PARFENOV$^2$, 
A.A.PAVLOV$^2$, D.P.PETUKHOV$^1$, P.G.POKHIL$^1$, P.A.POKOLEV$^2$, 
M.I.ROZANOV$^5$, V.YU.RUBZOV$^2$, I.A.SOKALSKI$^1$, CH.SPIERING$^8$, 
O.STREICHER$^8$, B.A.TARASHANSKY$^2$, T.THON$^8$, D.B.VOLKOV$^2$, 
CH.WIEBUSCH$^8$, R.WISCHNEWSKI$^8$\\[2mm]
}

\address{
1 - Institute  for  Nuclear  Research,  Russian  Academy  of   Sciences
(Moscow); \mbox{2 - Irkutsk} State University (Irkutsk); \mbox{3 - Moscow}
State University (Moscow); \mbox{4 - Nizhni}  Novgorod  State  Technical
University  (Nizhni   Novgorod); \mbox{5 - St.Petersburg} State  Marine
Technical  University  (St.Petersburg); \mbox{6 - Kurchatov} Institute
(Moscow); \mbox{7 - Joint} Institute for Nuclear Research (Dubna);
\mbox{8 - DESY} Institute for High Energy Physics (Zeuthen) 
}

\vspace{1cm}
\vspace{-12pt}
\maketitle\abstracts{    
The first stage of the Baikal Neutrino Telescope NT-200, the 
detector NT-36, was operated
from 1993 to 95. The data obtained with this small array
 were analysed to search 
for vertically upward muons. 
Apart from neutrinos
generated in the atmosphere at the opposite side of the Earth, such muons
might be due to neutrinos produced in neutralino annihilations in the
center of the Earth. We have selected two clear neutrino candidates.
From this, an 90\% upper limit of 1.3$\cdot$10$^{-13}$
muons cm$^{-2}$ sec$^{-1}$ in a cone with 15 degree half-aperture around the opposite
zenith is obtained for muons due to neutralino annihilation.
}

\setlength{\parindent}{1cm}
\vspace{1mm}
\section{Introduction}

The Baikal Neutrino Telescope NT-200 is being deployed in the Siberian Lake Baikal~\cite{1}.
In April 1993, 
its first stage,
the detector NT-36 
with 36 PMTs at 3 strings, was put into operation. 
Being slightly modified in March-April, 1994, it
took data up to March 1995. 
There were 6 PMT
pairs along each of the 3 strings of NT-36. 
The two PMTs of a pair are switched in coincidence and represent
a recording channel.
The orientation of the channels 
from top (channel \#1) to bottom (channel \#6) at each string was 
{\em down-up-down-up-down-up} for the period from April, 1993 till March, 1994 and
{\em up-down-up-down-up-down} for the period from April, 1994 till March, 1995.
The array NT-36 was the first underwater detector 
with the capability
to perform  full spatial track reconstruction.
Atmospheric muon angular distributions 
experimentally obtained with the standard reconstruction~\cite{3} and NT-36 data
are well described by MC expectations~\cite{4}.
However, due to small value of the $S/N$ ratio ($S/N\approx 1/50$, 
where $S$ is rate
of upward neutrino induced events and $N$ is rate of downward atmospheric muons which
are misreconstructed as upward events), it is impossible to observe a clear neutrino
signal with NT-36 data and the standard reconstruction procedure.

However, for neutrinos coming nearly straight upward, $S/N$ has to be much better - 
once due to the noise steeply falling with increasing zenith angle, secondly since
an up-down rejection may be achieved even in the case of no full track reconstruction,
i.e. for events hitting only channels from one or two strings, but applying criteria
tailored to this case. Here we discuss this method of neutrino event selection and
present results obtained with NT-36 data. 

\section{Method}

In contrast with our standard reconstruction strategy, which suppose $\geq$6 
hits at $\geq$3 strings (necessary for full spatial reconstruction), we did
not perform a reconstruction at all, but applied cuts, which effectively reject
all events with the exception of nearly vertically moving upward muons. We selected
events triggering $\geq$4 channels (3 looking down and at least one looking
up) exclusively along a single string, since the tracks of the objects searched for
have nearly the same vertical orientation as the strings.

We tested the following off-line selection criteria for selected events:

\vspace{-2mm}
\begin{description}
 \item[1.]
   Time differences between any two hit channels $i$ and $j$ must
obey the inequality

\vspace{-3mm}
\begin{center}
             $|(t_{i}-t_{j})-(T_{i}-T_{j})|<dt$
\end{center}

\vspace{-3mm}
where $t_i(t_j)$ are the measured times in channels $i(j)$, 
$T_i(T_j)$ are the ``theoretical'' times expected for minimal ionizing,
up-going vertical muons and $dt$ is a time cut. 

\vspace{-2mm}
 \item[2.]
      The minimum value of amplitude asymmetries for all pairs of alternatively
directed hit channels must obey the inequality

\vspace{-3mm}
\begin{center}        
$dA_{ij}(down-up) > 0.3$, 
\end{center}

\vspace{-3mm}
where $dA_{ij}(down-up)= (A_{i}(down)-A_{j}(up))/(A_{i}(down)+A_{j}(up))$ and 
$A_{i}(down)\linebreak 
(A_{j}(up))$ are the amplitudes of channel $i(j)$ facing downward(upward).

\vspace{-2mm}
 \item[3.]
    All amplitudes of downward looking hit channels must exceed 4 photoelectrons

\vspace{-3mm}
\begin{center}        
$A_i(down)>4 ph.el$.
\end{center}

\vspace{-3mm}
 \item[4.]
  The amplitude asymmetry   $dA(down-down)$
for downward looking hit channels is defined
as that of the 3 possible combinations 
$dA_{ij}(down-down)=(A_i-A_j)/(A_i+A_j)\mid _{i>j}$ with the
largest absolute value. For background events due to showers below the
array it peaks at values close to 1, for vertical neutrino candidates
it should be close to zero.
 The fourth criterion rejects half of the neutrino sample and
 nearly all events due to deep showers from downward atmospheric muons:

\vspace{-3mm}
\begin{center}        
            $dA(down-down)<0$.
\end{center}
\end{description}
\medskip

The dependence of the expected yearly number of muons generated 
by atmospheric neutrinos
and of background events
on the time cut $dt$  are presented in Fig.1.   
The 
curves marked 1, 2, 3 and 4 correspond to the 
trigger conditions {\it 1, 1-2,
1-3} and {\it 1-4}, respectively. The 'crosses' denote background 
curves and 'asterisks' denote 
muons from atmospheric neutrinos. 
One sees that the signal-to-noise ratio $S/N$
is close to 1 for trigger conditions {\it 1-3} and $dt\leq20ns$
and improves with decreasing $dt$ or applying criterion {\it4}.

\newpage

\strut

\vspace{1cm}

\begin{figure}[h] 
\centering
\mbox{\epsfig{file=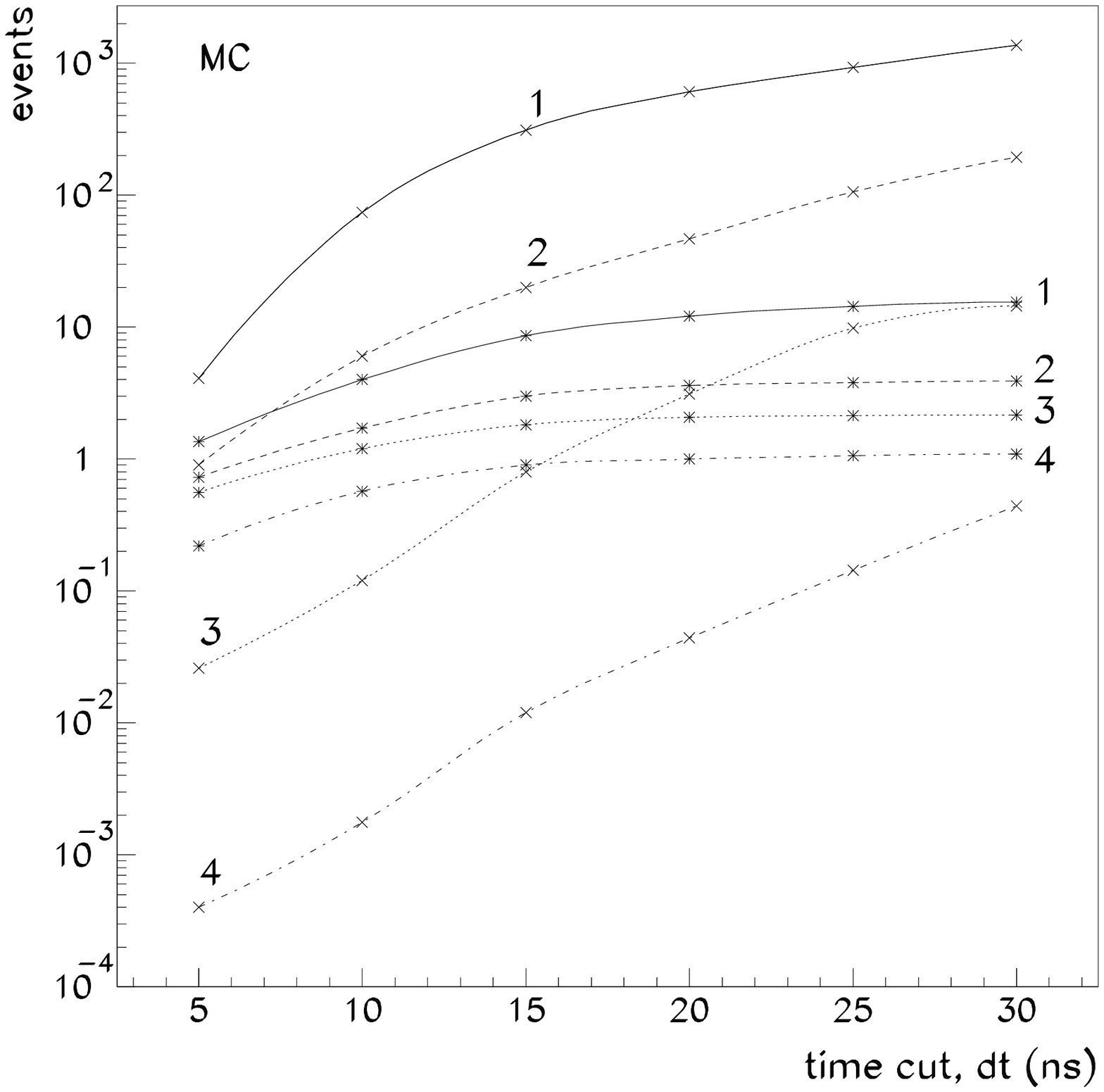,height=16cm}}
\end{figure}

\vspace{0cm}

{\bf Figure 1:} Expected numbers of muons from atmospheric neutrinos 
(asterisks) and background events (crosses) per year vs. time cut $dt$. 
Curves marked 1; 2; 3 and 4 correspond to trigger conditions {\it 1, 1-2, 1-3}
and \mbox{{\it 1-4.}}

\newpage

\strut

\vspace{1cm}

\begin{figure}[h] 
\centering
\mbox{\epsfig{file=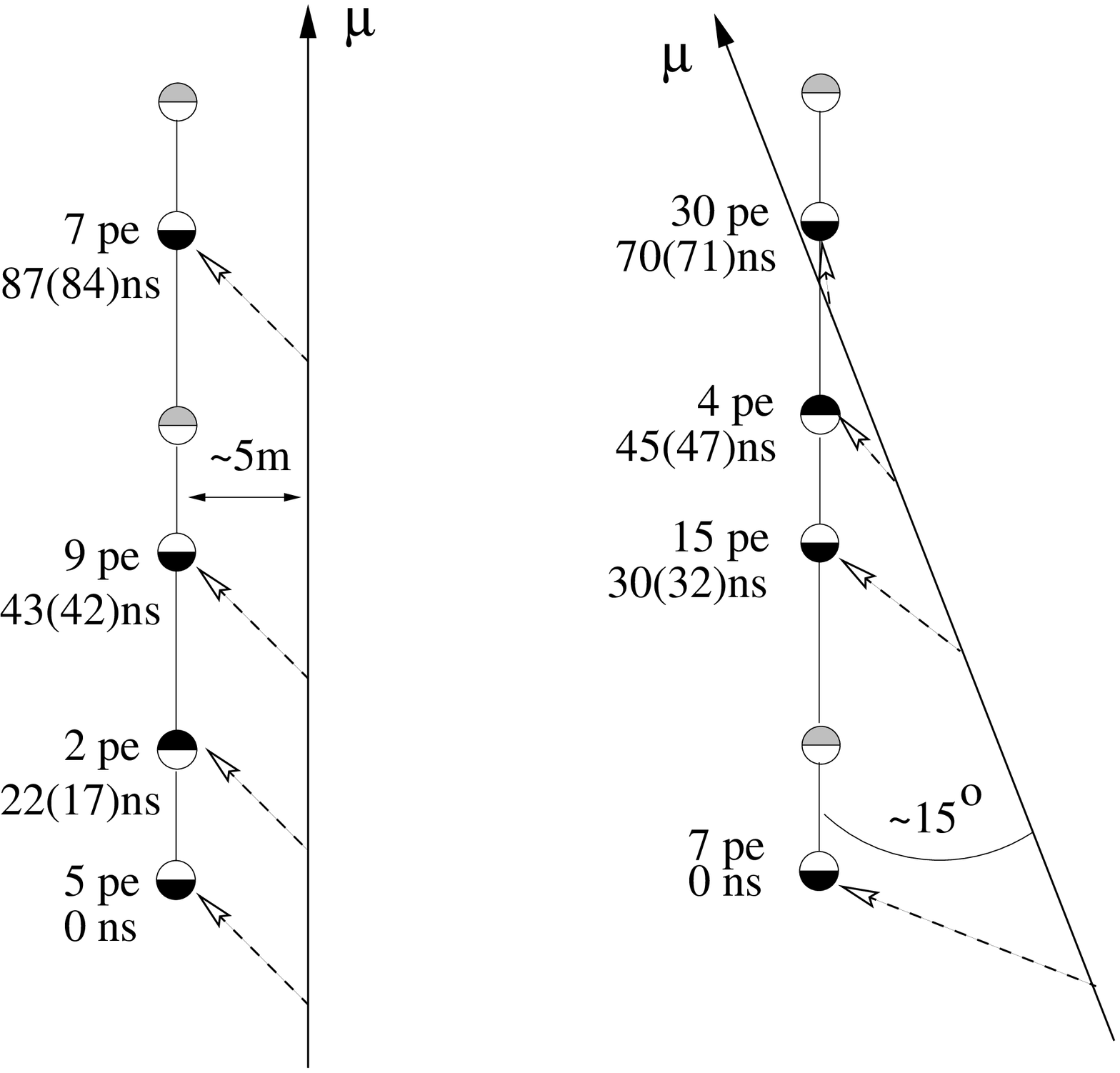,height=14cm,angle=-90}}
\end{figure}

\vspace{1cm}

{\bf Figure 2:} The two neutrino candidates. The hit PMT pairs (channels) are
marked in black. Numbers give the measured amplitudes (in photoelectrons) and 
times with respect to the first hit channel. Times in brackets are those 
expected for a vertical going upward muon (left) and an upward muon passing 
the string under \mbox{$15^o$} (right).

\newpage

\strut

\vspace{0cm}

\begin{figure}[h] 
\centering
\mbox{\epsfig{file=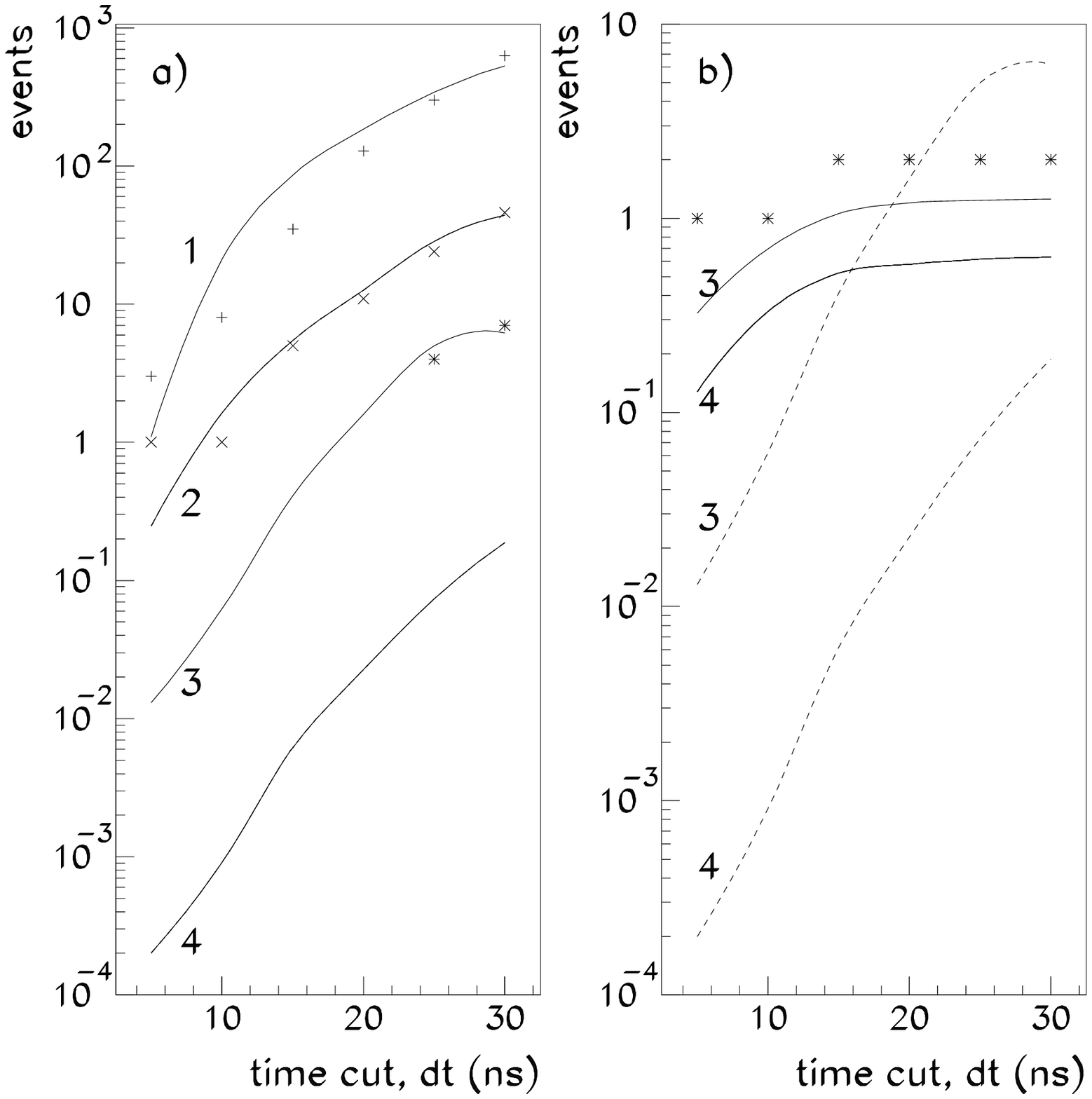,height=16cm}}
\end{figure}

\vspace{1cm}

{\bf Figure 3:} Distribution of experimental sample vs. time cut $dt$. 
Numbers 1; 2; 3 and 4 correspond to trigger conditions {\it 1, 1-2,1-3} and 
{\it 1-4}, respectively. a) - background events: lines present MC expectations 
for different trigger conditions; b) - neutrino candidates: solid and dashed 
lines present MC expectations for upward going  muons generated by atmospheric
neutrinos (not taking into account light scattering in water) and for 
background events.  

\newpage

\section{Results}

The  analysis presented here is based on the data taken with
\mbox{NT-36} between April 8, 1994
and March 5, 1995 (212 days lifetime).
Upward-going muon candidates were selected from a total of 
\mbox{$8.33\cdot 10^{7}$} events recorded 
by the muon-trigger \mbox{"$\geq 3$} hit channels". 
The samples fulfilling trigger conditions {\it 1, 1-2, 1-3} 
and {\it 1--4} with time cut $dt=20$ns contain 131, 17 and 2 events,
respectively. 
Only two events fulfill trigger conditions {\it 1--3} and {\it 1--4}. 
These events were recorded at 6 June and 3 July 1994.
The first event is consistent with a nearly vertical upward going muon
and the second one with an upward going muon with
zenith angle $\theta_{\mu}=15^{\circ}$ (Fig.2).

Fig. 3 shows the passing rate for two samples of events in dependence
on the
time cut $dt$. The "experimental neutrino sample" consists of just the
two events shown in Fig.2, the "experimental background sample"
contains all other events with the exception of these two.
MC curves have been obtained from modelling upward muons from
atmospheric neutrinos ("neutrinos") and from downward going
atmospheric muons ("background").

Fig. 3a demonstrates that MC describes the data within a factor of
3-4. From fig. 3b one sees that the probability to observe a
background event with $dt<20$nsec is about 2 percent only.
Whereas the shapes of  experimental and MC  distributions 
in Fig.3b are quite similar, the absolute values disagree by
a factor of 1.5-4, depending on the criterion.
The MC calculated numbers of upward going muons 
are systematically below the two experimentally observed events.
Apart from statistics, the reason may be the following: 
MC simulations of the NT-36 
response to upward going muons from
atmospheric neutrinos has been performed without taking into account
 light scattering in water.
A raw estimate shows that the expected number of detected 
upward going muons may rise by 40-80\%
when scattering process will be taken into account.  

Considering the two neutrino candidates as atmospheric neutrino 
events, a 90 \% CL upper limit
of $1.3 \cdot 10^{-13}$ 
(muons cm$^{-2}$ sec$^{-1}$) in a cone with 15 degree half-aperture around the opposite
zenith is obtained for upward going muons generated by neutrinos due to neutralino annihilation in the
center of the Earth. The limit corresponds to muons with energies greater than the threshold energy
$E_{th} \approx 6$ GeV, defined by  30m string length. This is still an order of magnitude higher
than the limits obtained by Kamiokande~\cite{5}, Baksan~\cite{6} and MACRO~\cite{7}.
The effective area of NT-36 for
nearly vertical upward going muons fulfilling our separation criteria {\it 1-3} with $dt=20ns$ is 
$S_{eff}=50$ m$^{2}$/string. A rough estimate of the effective area of
the full-scale 
Baikal Neutrino Telescope \mbox{NT-200} (with eight strings twice as long as those of
\mbox{NT-36}) with respect
to nearly vertically upward going
 muons gives $S_{eff} \approx 400-800$ m$^{2}$.

\end{document}